\documentclass[twocolumn]{aastex631}

\usepackage{mathtools}
\usepackage{nccmath}
\usepackage{threeparttablex} % or threeparttable if you're not using xparse
\usepackage{array}

\newrobustcmd{\B}{\bfseries}

\newcommand{\qfit}{\texttt{q3dfit}}
\newcommand{\pyqsofit}{\texttt{PyQSOFit}}
\newcommand{\pyneb}{\texttt{PyNeb}}

\newcommand{\oiii}{\hbox{[O$\,${\scriptsize III}]}}
\newcommand{\nii}{\hbox{[N$\,${\scriptsize II}]}}

\newcommand{\sii}{\hbox{[S$\,${\scriptsize II}]}}

\newcommand{\feii}{\hbox{[Fe$\,${\scriptsize II}]}}

\newcommand{\kms}{km\,s$^{-1}$} 
 
\newcommand{\eden}{cm$^{-3}$}

\renewcommand{\deg}{\ensuremath{^{\circ}}}

\shorttitle{F2M1106}
\shortauthors{Sankar et al.}
\graphicspath{{./}{figures/}}

\begin{document}
%\linenumbers
\title{First results from the JWST Early Release Science Program Q3D: AGN photoionization and shock ionization in a red quasar at z = 0.4}

%Outflow kinematics (paper 1)

\correspondingauthor{Swetha Sankar}
\email{ssanka10@jhu.edu}

\author[0000-0002-4419-8325]{Swetha Sankar}
\affiliation{Department of Physics and Astronomy, Bloomberg Center, Johns Hopkins University, 3400 N. Charles St., Baltimore, MD 21218, USA}

\author[0000-0001-6100-6869]{Nadia L. Zakamska}
\affiliation{Department of Physics and Astronomy, Bloomberg Center, Johns Hopkins University, 3400 N. Charles St., Baltimore, MD 21218, USA}

\author[0000-0002-1608-7564]{David S. N. Rupke}
\affiliation{Department of Physics, Rhodes College, Memphis, TN 38112, USA}

\author[0000-0003-3762-7344]{Weizhe Liu}
\affiliation{Department of Astronomy, Steward Observatory, University of Arizona, Tucson, AZ 85719, USA}

\author[0000-0003-2212-6045]{Dominika Wylezalek}
\affiliation{Zentrum für Astronomie der Universität Heidelberg, Astronomisches Rechen-Institut, Mönchhofstr 12-14, D-69120 Heidelberg, Germany}

\author[0000-0002-3158-6820]{Sylvain Veilleux}
\affiliation{Department of Astronomy and Joint Space-Science Institute, University of Maryland, College Park, MD 20742, USA}

\author[0000-0002-6948-1485]{Caroline Bertemes}
\affiliation{Zentrum für Astronomie der Universität Heidelberg, Astronomisches Rechen-Institut, Mönchhofstr 12-14, D-69120 Heidelberg, Germany}

\author[0009-0003-5128-2159]{Nadiia Diachenko}
\affiliation{Department of Physics and Astronomy, Bloomberg Center, Johns Hopkins University, 3400 N. Charles St., Baltimore, MD 21218, USA}

\author[0000-0002-9932-1298]{Yu-Ching Chen}
\affiliation{Department of Physics and Astronomy, Bloomberg Center, Johns Hopkins University, 3400 N. Charles St., Baltimore, MD 21218, USA}

\author[0000-0001-7572-5231]{Yuzo Ishikawa}
\affiliation{MIT Kavli Institute for Astrophysics and Space Research, Massachusetts Institute of Technology, Cambridge, MA 02139, USA}

\author[0000-0002-0710-3729]{Andrey Vayner}
\affiliation{Department of Physics and Astronomy, Bloomberg Center, Johns Hopkins University, 3400 N. Charles St., Baltimore, MD 21218, USA}

\author[0000-0001-5783-6544]{Nicole P. H. Nesvadba}
\affiliation{Université de la Côte d'Azur, Observatoire de la Côte d'Azur, CNRS, Laboratoire Lagrange, Bd de l'Observatoire, CS 34229, Nice cedex 4 F-06304, France}

\author[0000-0003-4286-5187]{Guilin Liu}
\affiliation{CAS Key Laboratory for Research in Galaxies and Cosmology, Department of Astronomy, University of Science and Technology of China, Hefei, Anhui 230026, China}
\affiliation{School of Astronomy and Space Science, University of Science and Technology of China, Hefei 230026, China}

\author[0000-0003-4700-663X]{Andy D. Goulding}
\affiliation{Department of Astrophysical Sciences, Princeton University, 4 Ivy Lane, Princeton, NJ 08544, USA}

\author[0000-0003-0291-9582]{Dieter Lutz}
\affiliation{Max-Planck-Institut für Extraterrestrische Physik, Giessenbachstrasse 1, D-85748 Garching, Germany}

\begin{abstract}
Red quasars, often associated with potent \oiii\ outflows on both galactic and circumgalactic scales, may play a pivotal role in galactic evolution and black hole feedback. In this work, we explore the \feii\ emission in one such quasar at $z = 0.4352$—F2M J110648.32+480712.3—using the integral field unit (IFU) mode of the Near Infrared Spectrograph (NIRSpec) aboard the James Webb Space Telescope (JWST). Our observations reveal clumpy \feii\ gas located to the south of the quasar. By comparing the kinematics of \feii\ and \oiii, we find that the clumpy \feii\ gas in the southeast and southwest aligns with the outflow, exhibiting similar median velocities up to v$_{50}$ $=$ 1200 \kms\ and high velocity widths W$_{80}$ $>$ 1000 \kms. In contrast, the \feii\ gas to the south shows kinematics inconsistent with the outflow, with W$_{80}$$\sim$ 500 \kms, significantly smaller than the \oiii\ at the same location, suggesting that the \feii\ may be confined within the host galaxy. Utilizing standard emission-line diagnostic ratios, we map the ionization sources of the gas. According to the MAPPINGS III shock models for $\feii$/Pa$\beta$, the regions to the southwest and southeast of the quasar are primarily photoionized. Conversely, the \feii\ emission to the south is likely excited by shocks generated by the back-pressure of the outflow on the galaxy disk, a direct signature of the impact of the quasar on its host.

\end{abstract}

%% Do not enter keywords now, they are entered at submission

\section{Introduction} 
\label{sec:intro}

One crucial aspect of quasar feedback is the interplay between quasar-driven winds/outflows and the interstellar medium (ISM) within the host galaxy \citep{fabi12}. These powerful winds impart a substantial amount of energy onto the surrounding ISM, leading to the formation of shock waves that can have profound consequences for the galaxy's star formation capacity, either by dispersing or reheating its cold gas reservoir \citep{silk98, king03, korm13,king15, vei2020}. Consequently, the identification and characterization of shock signatures have long been sought after as a diagnostic tool for understanding the interaction between quasar winds and the ISM. 

Historically, evidence for this feedback between quasar driven winds/outflows and shock-heated regions has been primarily indirect and observed in lower luminosity active galactic nuclei (AGN) residing in nearby galaxies, as illustrated in \citet{Allen99}. For example, \citet{zaka14} use weaker optical lines as indirect shock signatures in a set of obscured luminous quasars, interpreting the similar velocities between these emissions and the outflows as evidence of quasar-driven winds propagating into the ISM of the host galaxy. The advent of integral field spectroscopy has ushered in an era marked by the accumulation of direct, spatially resolved evidence for shock signatures beyond the quasar's illumination cone. This is seen in \citet{riff21}, who provide direct evidence of outflow driven shock ionization in regions orthogonal to the ionization axis in Mrk79 and Mrk348. Similarly, \citet{Leung21} discovered a region of low-ionization, high dispersion gas in Mrk273, displaying an orthogonal orientation to the photoionization cone. These findings have also been extended beyond the local universe more recently, as \citet{vayner23a} find evidence for shock ionization induced by a quasar-driven outflow in a z = 3 redshift source, J1652, through JWST NIRSpec data. 

The low-ionization, near-infrared emission lines of \feii\ are strong in regions of fast shocks, effectively tracing warm and cold interstellar medium phases \citep{oliva1989}. These lines frequently serve as tracers to identify shocked regions. However, unraveling the origin of \feii\ is complicated. While prior studies suggest quasar photoionization either directly or through quasar-driven outflows to dominate ionization due to high temperatures of the ionized regions \citep{mour00}, other mechanisms could potentially affect the interpretation of \feii\ in these regions. For instance, several studies find \feii\ emission to arise from star formation processes, where supernovae release vast amounts of energy, launching powerful shockwaves throughout the ISM \citep{reach2002, reach2005, koo2007, lee2009}. The increased gas phase abundance of iron from dust grain destruction combined with other ionizing effects is responsible for large \feii\ enhancements as observed in several supernovae. Additionally, hydrodynamical simulations \citep{muk2016, wagner2016} and observational studies \citep{lacy2017, zovaro2019} cite radio jets as potential \feii\ production mechanisms. These jets propagate into inhomogeneous ISM to form energy-driven bubbles that ionize the surrounding medium and create bow shocks. 

In star forming galaxies, the emission of \feii\ exhibits a correlation with the star formation rate, enabling differentiation between contributions from quasar photoionization and those originating from star formation processes \citep{gra1987, koolee2015}. If the detected \feii\ emission is in excess of what is expected based on star formation alone, it can be attributed to the influence of the quasar.

We obtained integral field spectroscopy data of \feii\ with the James Webb NIRSpec instrument as part of the Q3D Early Release Science program. The Q3D program is dedicated to the exploration of quasar feedback processes in host galaxies and development of a scientific tool, \qfit\ \citep{q3dfit23}. This software provides the scientific community with more convenient spectral mapping and point spread function (PSF) decomposition of the JWST IFU data to obtain science-ready products that can be used for the analysis of quasars and their galaxy hosts on different redshifts. F2M J110648.32+480712.3, hereafter F2M1106, is one of three quasars we use to train and test the \qfit\ package. 

In this paper, we begin by summarizing all known properties of F2M1106 in Section \ref{sec:obj}. Section \ref{sec:obs} details the observational setup, data reduction, and analysis of the \feii\ emission in the JWST NIRSpec dataset. In Section \ref{sec:discussion}, we discuss the origin of \feii\ ionization. We summarize in Section \ref{sec:conclusions}. Throughout this paper we adopt the cosmological parameters H$_{0}$ = 69.6 km s$^{-1}$ Mpc$^{-1}$, $\Omega_{M}$ = 0.286, and $\Omega_{\Lambda}$ = 0.714 of the flat $\Lambda$CDM model \citep{bennett2014}.

\section{F2M1106} 
\label{sec:obj}

\subsection{Previous Studies} \label{subsec:props}

F2M1106 is a red quasar originally identified by \citet{glik12} and confirmed by them using optical spectroscopy. It has a redshift of $z = 0.4352$ as determined by Keck KCWI based on stellar kinematics (Rupke et al., in prep). The initial identification of the object was based on its detection as a luminous radio source in the Faint Images of the Radio Sky at Twenty-Centimeters survey (FIRST; \citealt{whit97}) indicating the presence of an active nucleus and red colors in the Two Micron All Sky Survey (\citealt{skru06}) suggesting high levels of reddening. A broader coverage optical spectrum of the source was further obtained by the Baryon Oscillation Spectroscopic Survey (BOSS) of the Sloan Digital Sky Survey (SDSS; \citealt{eise11}). It shows broad emission lines of H$\alpha$, H$\beta$ and H$\gamma$, a strong contribution from FeII \citep{boro92} which was analyzed by \citet{hu08} as part of a larger sample and flat continuum ($F_{\lambda}\propto \lambda^0$, unlike the optical continuum of unobscured quasars with $F_{\lambda}\propto \lambda^{-1.56}$, \citealt{vand01}).

The quasar is not detected in a 6 ksec {\it Chandra} exposure \citep{glik24}, which is not unusual for red and obscured quasars, even intrinsically luminous, due to the high levels of X-ray absorption \citep{lama16, glik17}. The radio source, with flux $F_{\nu}=10.6\pm0.14$ mJy at 1.4 GHz \citep{whit97}, is consistent with being point-like at the 5\arcsec\ spatial resolution of FIRST. The flux at 150 MHz obtained in LOFAR Two-metre Sky Survey \citep{shim17} is $F_{\nu}=75\pm 15$ mJy, suggesting an overall spectral index of $\alpha=-0.88$, standard for optically thin synchrotron radiation. There is no evidence for a strong radio core emission which would be indicative of a compact jet.

\citet{shen23} obtained optical integral field spectroscopy of the object with the Gemini Multi-Object Spectrograph and used spatially resolved kinematic mapping of \oiii$\lambda$5007 \AA\ to discover a large ($\sim40$ kpc from end to end) ionized nebula around the target. The gas shows an organized motion with a velocity difference between the redshifted and the blueshifted sides of over 1000 km s$^{-1}$. \citet{shen23} interpret these data as a powerful quasar-driven galactic outflow, possibly in the form of `super-bubbles' \citep{gree12} which are expected to occur when a wind breaks out of a dense environment of the host galaxy and into the lower density circumgalactic halo, with the galaxy's morphology influencing how the superbubbles evolve. Observations with {\it HST} reveal that low-redshift red quasars as a population are typically found in major mergers \citep{urru08}; however, no {\it HST} data are available for our target. While the host galaxy was not detected either in the integrated optical spectrum or in the Gemini spatially resolved spectroscopy, it is detected in the blue part of the optical range using deep Keck Cosmic Web Imager integral field spectroscopy and spectral PSF subtraction (Rupke et al., in prep).

Integral field spectroscopy of F2M1106 using the Mid-InfraRed Instrument (MIRI) on {\it JWST} detected the \oiii$\lambda$5007 \AA\ outflow in the light of [\ion{S}{4}]~10.51\,$\mu$m \citep{rupk23}. This transition traces similar physical conditions to \oiii$\lambda$5007 \AA, but at much longer wavelengths. The comparison reveals differential dust reddening in the approaching and receding sides of the outflow, suggesting that the latter is obscured by the host galaxy at $A_V \leq 2.0$~mag, discussed further in Section \ref{sec:extinction}. 

\subsection{Spectral Energy Distribution}
\label{subsec:sed_sec}

\begin{figure}
    \centering
    \includegraphics[width=0.45\textwidth]{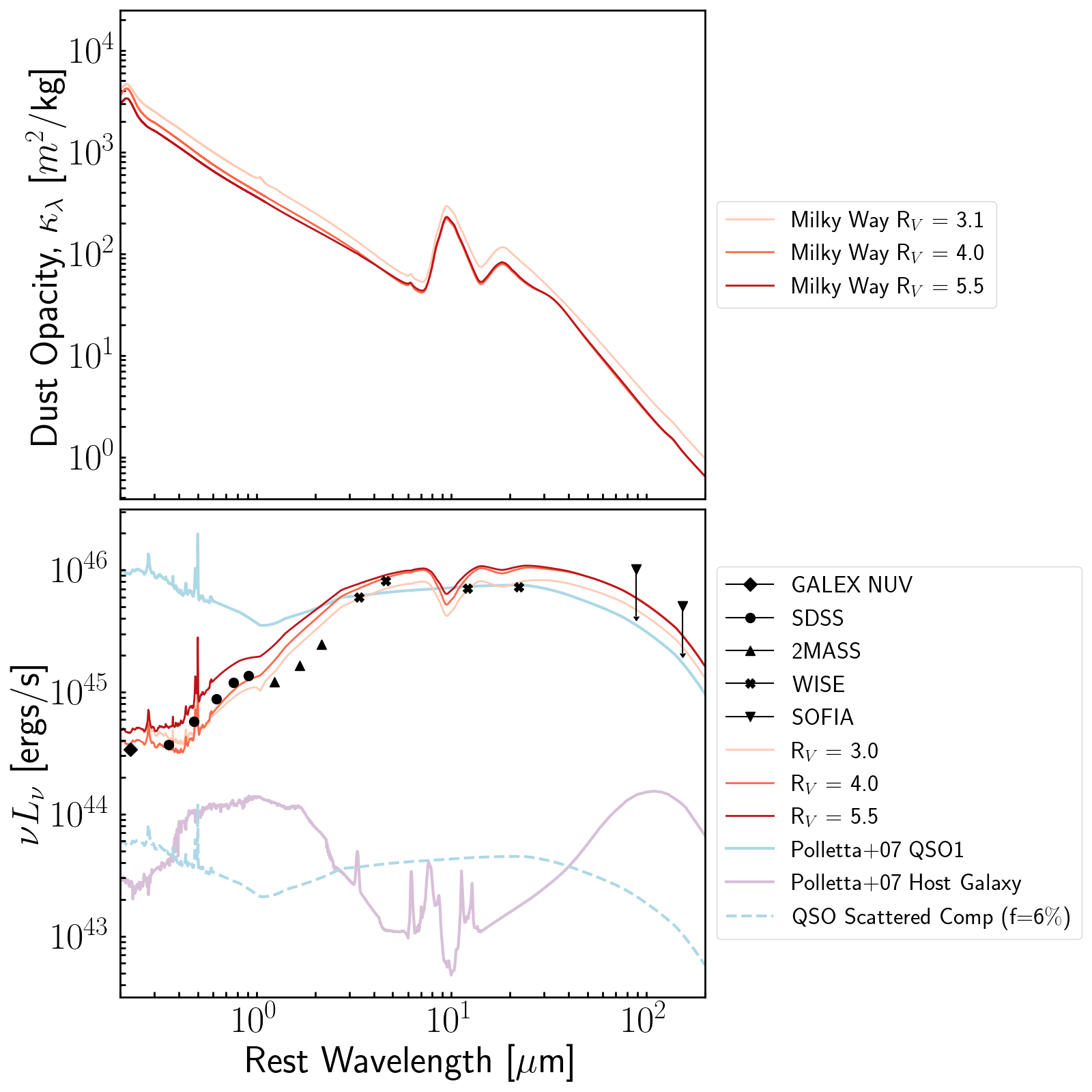}\\
     \caption{Top: Dust opacity curves \citep{wein01}, each with a different Milky Way dust extinction coefficient. Bottom: Our spectral energy distribution (SED) models for F2M1106 multiwavelength photometry (error bars are present but small). The intrinsic AGN continuum experiences considerable reddening with observed UV/ optical excess in the SDSS u band and GALEX NUV. We utilize a model consisting of a blue quasar \citep{poll07} (in blue), a dusty torus (not pictured; \citet{wein01}), and with contributions from the host galaxy (purple) and quasar scattered light (dashed blue). Three fits of this model are depicted (in various shades of red), each corresponding to the three distinct dust models illustrated in the upper plot. We find the scattered light component to comprise roughly 6$\%$ of the intrinsic UV emission.}
     \label{fig:sed}
    
\end{figure}

We model the spectral energy distribution (SED) spanning from the optical to far-infrared wavelengths. The SED incorporates photometric data obtained from various surveys and instruments, including the Galaxy Evolution Explorer (GALEX) near-ultraviolet (NUV) band \citep{martin2005}, SDSS {\em ugriz} bands in the optical, 2MASS J-, H-, K-bands, and Wide-field Infrared Survey Explorer (WISE) W1--W4 channels \citep{wrig10}. Additionally, we obtain 3$\sigma$ upper limits in the far-infrared from Stratospheric Observatory for Infrared Astronomy (SOFIA) observations using the HAWC+ C (62 \micron\ rest-frame) and D (107 \micron\ rest-frame) bands \citep{harp18}, which are observed for 67 min and 77 min respectively.

Red quasars, typically exhibiting high luminosities (i.e $\nu$L$_{5\micron}$ = 10$^{45.5}$ erg s$^{-1}$) and reddening ($A_V\sim 1-20$ mag; \citealt{glik07, urru12, Ishikawa21, glik22}), outshine their host galaxies in the rest-frame optical. Their SEDs are often well characterized by reddening an unobscured quasar template spectrum using a `screen of cold dust' approximation with a standard dust extinction curve. We find similarly high WISE luminosities for our object and initially consider a model with a blue quasar \citep{poll07} subject to varying levels of dust absorption \citep{wein01}. However, the fitting proves inadequate in explaining the observed blue excess, particularly prominent in the SDSS u and GALEX NUV bands. In the rest-frame UV, the quasar light is heavily extinguished and the host galaxy, particularly the young stars within it, may contribute to the overall flux density \citep{jahn04}. Additionally, the UV-emitting component could arise from a fraction of quasar light scattering off of the ISM on scales larger than the obscuring material \citep{zaka05, zaka06}.

%From Keck KCWI observations, we know the host galaxy flux to be in the range 16-33$\%$ of the total quasar light over the rest frame range 2400--3900 \AA\ (Rupke et al. in prep.). We scale the template of an active star forming galaxy \citep{poll07} assuming the host contributes 25$\%$ of the total quasar light in this range (the average) and compute a star formation rate upper limit for the host of $\sim$132 M$_{\odot}$ yr$^{-1}$ using the 8--1000 \micron\ infrared luminosity \citep{ebell03}. 

The stellar component of the host galaxy is directly detected in Keck KCWI observations. From these data, we estimate that the host galaxy contributes 16–33$\%$ of the total quasar light over the rest-frame wavelength range 2400–3900 \AA\ (Rupke et al., in prep.). To account for this contribution, we incorporate an active star-forming galaxy template from \citet{poll07} into our analysis. We scale the template to represent 25$\%$ of the total quasar light in this range (the average contribution) and use it to estimate a star formation rate upper limit for the host galaxy. This yields an upper limit of $\sim$ 132 M$_{\odot}$ yr$^{-1}$ using the 8--1000 \micron\ infrared luminosity \citep{ebell03}. Because there is a wide range of infrared-to-UV flux ratios amongst star-forming galaxy templates and because we do not have good far-infrared / submm coverage of the SED, this is only an estimate. 

The remaining UV contribution is accounted for by a blue quasar component scattered into our line of sight with efficiency f$_{scat}$$=$6$\%$, which is consistent with previous studies who generally quote efficiencies of a few per cent. For example, \citet{youn09} find a scattering efficiency of 6.25$\%$ for NGC 1068 while \citet{gree14a} find an efficiency of 3$\%$ for radio quiet quasar SDSS J135646.10+102609.0. \citet{obie16} analyze Hubble Space Telescope observations of 20 luminous obscured quasars between 0.24 $<$ z $<$ 0.65 and find a median scattering efficiency of 2.3$\%$.

The resulting model, which includes a blue quasar, cold screen of dust modeled using three different dust opacity curves with varying R$_{V}$, a star forming galaxy, and a quasar scattered component, is illustrated in Figure \ref{fig:sed}. We do not adopt a single best-fit dust opacity curve but instead present results for all three modeled SED fits. Our analysis yields a moderate continuum extinction magnitude, A$_{V, SED}$ $=$ 3.2 $\pm$ 0.8 mag. Estimating the bolometric luminosity using the wavelength-dependent bolometric correction to the monochromatic luminosity at 3000 \AA for Type 1 quasars \citep{maca04, rich06}, we derive a value of L$_{bol}$ $=$ 10$^{46.5 \pm 0.1}$ ergs s$^{-1}$. We choose not to perform a more sophisticated multi-parametric SED fit because the poor coverage of the SED in the far-infrared precludes an independent measurement of the SFR, because the scattered light contribution and the UV-bright star formation are degenerate \citep{gree24}, and because in red quasars specifically the standard packages tend to result in an overly massive fit to the host galaxy as they do not reproduce well the near-infrared portion of the quasar emission.

\subsection{Optical Integrated Spectrum Analysis}

%introduce pyqsofit and fitting procedure.
We employ the \pyqsofit\footnote{\url{https://github.com/legolason/PyQSOFit}} code (v2.0) developed by \citet{guo2018} to fit the optical nuclear spectrum obtained from SDSS. The fitting procedure is conducted in the rest frame using a $\chi^{2}$-based methodology. The continuum is modeled using a third-order polynomial function. During the continuum fitting, all major emission line regions are masked to prevent contamination of the continuum model by line emission, with the continuum fit performed over the remaining spectral regions. A UV-optical FeII emission template from \citet{boro92} is simultaneously fitted and subtracted along with the continuum. We then proceed to model the emission lines, including H$\alpha$ and H$\beta$, utilizing a combination of narrow and broad Gaussian components. Within the H$\alpha$+\nii+\sii\ complex, all forbidden lines and the narrow component of H$\alpha$ have the same velocity centroids and widths. Broad components like H$\alpha$ and H$\beta$, which originate from the broad-line region, vary independently. For accurate flux scaling, fixed flux ratios are enforced where appropriate, such as for \oiii$\lambda$4959 \AA\ and for \oiii$\lambda$5007 \AA\ and \nii$\lambda$6548 \AA\ and \nii$\lambda$6584 \AA. 

\begin{figure*}
    \centering
    \includegraphics[width=0.9\textwidth]{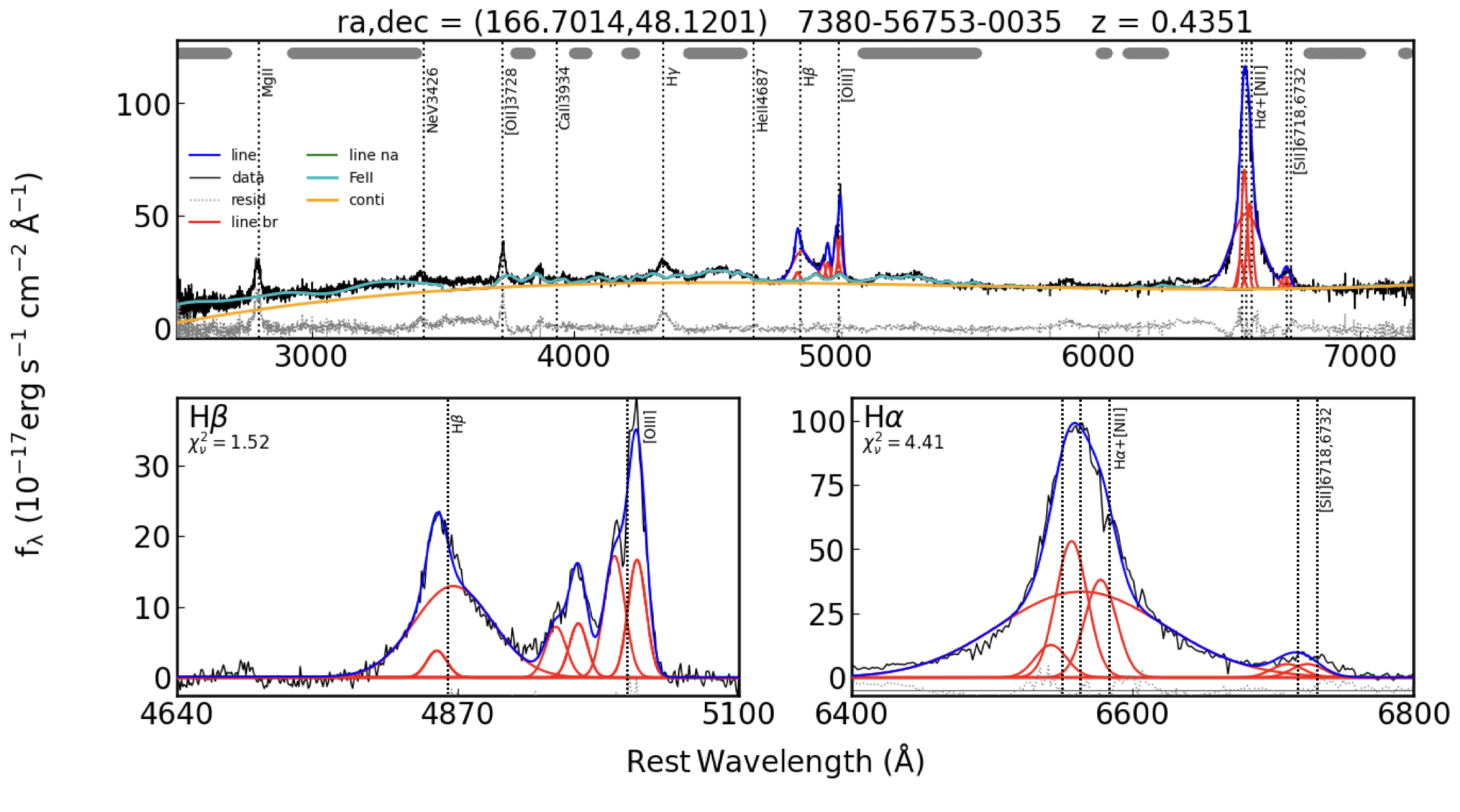}\\
    \caption{\pyqsofit\ fits to optical Sloan Digital Sky Survey spectra of F2M1106. The upper panel shows the fits to the entire spectrum. The continuum is shown in orange, the optical FeII contribution is in aqua. In the bottom panels: \textit{(left)} the H$\beta$, \oiii\ $\lambda$4959, and \oiii\ $\lambda$5007 emission features. The \oiii\ lines require two Gaussian components during fitting while H$\beta$ favors an unshifted broad component and a blueshifted broad component. \textit{(right)} H$\alpha$ fit with two components. Other lines fit here are \nii\ $\lambda$6549, \nii\ $\lambda$6585, \sii\ $\lambda$6718, and \sii\ $\lambda$6732. Residuals are shown in each plot in dotted gray with a vertical offset for visual clarity.}
    \label{fig:sloanspec}
\end{figure*}

The fitting results are shown in Figure \ref{fig:sloanspec} and the best-fit parameters per emission line is given in Table \ref{tab:pyqsofit_lines}, with the uncertainties determined through a Monte Carlo procedure over 200 realizations. While the model provides a good overall match to the observed spectrum, some limitations remain in fitting specific components. In particular, the shifted component of H$\beta$ and the redshifted components of the \oiii\ doublet appear slightly under-represented in intensity, causing a minor underestimation of the \oiii\ peak. We investigate increasing the intensity of these components to better match the observed peaks; however, this leads to a poorer overall fit. This limitation likely arises from the relatively simple model adopted, which uses a minimal number of Gaussian components to avoid overfitting. Although a more complex model might yield a closer match, it also introduces additional free parameters without strong justification from the data, and thus we refrain from this approach.

%Following the modeling of the continuum through a third order polynomial function and the subsequent subtraction of the continuum emission and an optical FeII template included in the package,

[O$\,${\scriptsize III}] emission is characterized by unusually high velocity width, as appropriate for a quasar with an extremely high-velocity circumgalactic outflow \citep{shen23}. Our analysis reveals that H$\alpha$ and H$\beta$ lines are significantly broader than any of the forbidden lines in the spectrum,  clearly indicating their origin in the broad-line region and confirming F2M1106's classification as a Type 1 AGN as stated in \citet{shen23}. Additionally, we find that the broad wings of H$\beta$ partially overlap with the \oiii\ doublet due to the significant velocity broadening of both species, making it more difficult to separate individual components in this region. We explore the possibility of these lines sharing similar kinematic structures, potentially indicating emission from the same physical region. However, we observe distinct kinematic signatures; the \oiii\ doublet exhibits a double-peaked, redshifted and blueshifted structure, contrasting with H$\beta$, which predominantly displays an unshifted broad component alongside a weaker and narrower blueshifted component. F2M1106 also has strong FeII emission \citep{hu08}, indicating that it may be a high-Eddington source \citep{boro92}. 

%we find the \oiii\ doublet to be blended together with H$\beta$ 

\subsection{Nuclear Extinction and Black Hole Mass Estimation}

% calculated quantities
We determine the extinction in the nuclear region in this object using the Balmer emission line ratio (using the broad H$\alpha$ and H$\beta$ components derived from fitting), which is shaped by the equilibrium between photoionization and recombination. Case B recombination in hydrogen sets a lower theoretical bound on H$\alpha$/H$\beta$ at 2.98 in dust-free gas at T$_{e}$ $=$ 10$^{4}$ K \citep{dopi2003}. In unobscured Type 1 quasars, this ratio peaks at 3.3 \citep{kim06}, and an elevated ratio beyond this suggests dust extinction. We estimate A$_{V}$ using the relation given in \citet{rif2021} to be 2.4 $\pm$ 0.3 mag, which is consistent with the extinction derived from the SED-fitting technique and with observations in other red quasars \citep{ban2015, faw2023}. 

We assume that the gas within the broad-line region surrounding the black hole is virialized \citep{pet2004} to estimate the black hole mass. The velocity of the gas is deduced from the width of the broad H$\alpha$ emission line, which is less affected than other tracers such as H$\beta$ by dust extinction, while the associated rest-frame line luminosity L$_{5100}$, which we correct for the extinction, serves as a proxy for the size of the broad-line region. Using the virial scaling relation from \citet{shenliu2012}, we derive a black hole mass of M$_{BH}$ = 10$^{9.5 \pm 0.2}$ M$_{\odot}$. The uncertainty is based solely on measurement errors from the MCMC-produced H$\alpha$ broad line FWHM, and does not account for potential systematic uncertainties. The black hole mass estimate could vary by an order of magnitude depending on the calibration methods employed for the emission lines \citep{bert2024}. We find this result to be similar to the black hole masses quoted for this object in \citet{hu08} and Rupke et. al. in prep, who determine the mass to be $\sim$10$^{9.1}$ M$_{\odot}$ using H$\beta$ and MgII respectively. The resulting Eddington luminosity is L$_{Edd}$ $=$ 10$^{47.5 \pm 0.8}$ erg s$^{-1}$ and the Eddington ratio, using the bolometric luminosity calculated in Section (\ref{subsec:sed_sec}), is log $\lambda_{Edd}$ $=$ --1.0$\pm$ 0.3. Table \ref{table:params} lists the properties in the 

\iffalse
\begin{table*}[htbp]
\centering
\caption{\pyqsofit\ Best-fit Emission Line Parameters}
\begin{tabular}{lllll}
\hline
\hline
Line ID & Component & $\lambda_{c}$ & Amplitude  & $\sigma$  \\
& & [\AA] & [$10^{-17}$ erg s$^{-1}$ cm$^{-2}$ \AA$^{-1}$] & [km s$^{-1}$] \\
\hline
H$\beta$        & Broad     & 4866 & 13.11 & 1950 \\
H$\beta$        & Narrow    & 4853 & 3.73  & 510  \\
{[OIII]}$\lambda$4959    & Wing      & 4969 & 7.67  & 450  \\
{[OIII]}$\lambda$4959    & Wing      & 4969 & 7.67  & 450  \\
{[OIII]}$\lambda$5007    & Wing      & 5017 & 16.65 & 450  \\
{[OIII]}$\lambda$5007    & Wing      & 5017 & 16.65 & 450  \\
H$\alpha$       & Broad     & 6563 & 33.64 & 2670 \\
H$\alpha$       & Narrow    & 6556 & 12.56 & 510  \\
{[NII]}$\lambda$6549     & Narrow    & 6542 & 5.16 & 510  \\
{[NII]}$\lambda$6585     & Narrow    & 6577 & 37.68  & 510  \\
{[SII]}$\lambda$6718     & Narrow    & 6710 & 5.16  & 510  \\
{[SII]}$\lambda$6732     & Narrow    & 6724 & 5.16  & 510  \\
\hline
\end{tabular}
\label{tab:pyqsofit_lines}
\end{table*}
\fi

%double check Halpha complex... something wrong about amplitudes

\begin{table*}[htbp]
\centering
\caption{\pyqsofit\ Best-fit Emission Line Parameters}
\begin{tabular}{lllll}
\hline
\hline
Line ID & Component & $\lambda_{c}$ & Amplitude  & $\sigma$  \\
& & [\AA] & [$10^{-17}$ erg s$^{-1}$ cm$^{-2}$ \AA$^{-1}$] & [km s$^{-1}$] \\
\hline
 H$\beta$        &  Broad     & \B 4866 $\pm$ \B 1 & \B 13.1 $\pm$ \B 1.3 & \B 1950 $\pm$ \B 300 \\
H$\beta$        & Narrow    & \B 4853.1 $\pm$ \B 0.5 & \B 3.7 $\pm$ \B 0.4 & \B 510 $\pm$ \B 27 \\
{[OIII]}$\lambda$4959    & Wing      & \B 4967.3 $\pm$ \B 0.2 & \B 7.7 $\pm$ \B 0.8 & \B 448 $\pm$ \B 11 \\
{[OIII]}$\lambda$4959    & Wing      & \B 4967.3 $\pm$ \B 0.2 & \B 7.7 $\pm$ \B 0.8 & \B 448 $\pm$ \B 11 \\
{[OIII]}$\lambda$5007    & Wing      & \B 5015.3 $\pm$ \B 0.2 & \B 16.7 $\pm$ \B 1.7 & \B 448 $\pm$ \B 11 \\
{[OIII]}$\lambda$5007    & Wing      & \B 5015.3 $\pm$ \B 0.2 & \B 16.7 $\pm$ \B 1.7 & \B 448 $\pm$ \B 11 \\
H$\alpha$       & Broad     & \B 6563 $\pm$ \B 2 & \B 30.5 $\pm$ \B 3.7 & \B 2670 $\pm$ \B 400 \\
H$\alpha$       & Narrow    & \B 6555.3 $\pm$ \B 0.5 & \B 47.7 $\pm$ \B 3.8 & \B 510 $\pm$ \B 30 \\
{[NII]}$\lambda$6549     & Narrow    & \B 6541.3 $\pm$ \B 0.5 & \B 12.2 $\pm$ \B 1.3 & \B 510 $\pm$ \B 30 \\
{[NII]}$\lambda$6585     & Narrow    & \B 6577.0 $\pm$ \B 0.5 & \B 33.6 $\pm$ \B 0.5 & \B 510 $\pm$ \B 30 \\
{[SII]}$\lambda$6718     & Narrow    & \B 6710.3 $\pm$ \B 0.5 & \B 5.2 $\pm$ \B 0.5 & \B 510 $\pm$ \B 30 \\
{[SII]}$\lambda$6732     & Narrow    & \B 6724.1 $\pm$ \B 0.5 & \B 5.2 $\pm$ \B 0.5 & \B 510 $\pm$ \B 30 \\

\hline
\end{tabular}
\label{tab:pyqsofit_lines}
\end{table*}

%$37.7 \pm 3.8$

\section{Analysis of JWST Data}
\label{sec:obs}
\subsection{Observational Design and Data Reduction}

James Webb Space Telescope NIRSpec integral field spectroscopy \citep{jakobsen22} observations were acquired on November 13, 2022 using both the G395H grating/ 290LP filter and the G235H grating/ 170LP filter. We employed a nine-point dither pattern during observation to improve the spatial sampling of the PSF. An additional exposure at the first dither position with the micro-shutter assembly (MSA) closed was taken to constrain any light contamination from bright objects in the instrument's field of view and remove it in the case of failed open shutters. Per detector, the effective exposure time per integration for the source was 233.422 s, with a total on-source exposure time of 2100.798 s.

We reduced the data using Space Telescope's JWST pipeline version 1.14.0\footnote{\url{https://github.com/spacetelescope/jwst}} in conjunction with the jwst\_1223.pmap version of the calibration reference files. In the initial stage, which encompasses standard infrared detector processing of the uncalibrated files, the pipeline subtracts the dark current, flags the data quality, and conducts an initial iteration of cosmic ray removal. Prior to advancing to the subsequent stage, we apply a band correction to mitigate significant instrument-induced variations across detectors and grating/filter combinations. The resulting rate files obtained from this stage were then fed into the second stage of the pipeline.

In the second stage, the pipeline assigns each frame a world coordinate system, followed by background subtraction, flat-fielding, and absolute flux calibration. Additionally, this stage converted the 2D spectra into a 3D data cube through the execution of the ``cube build" routine. We also identified and accounted for imprints generated by the open NIRSpec micro-shutters, as well as flagged and excluded any defective pixels. Finally, we execute the last stage of the pipeline to combine the exposures obtained from different dither positions utilizing the 3D emsm algorithm. We applied a final three sigma clipping method to remove residual cosmic rays, resulting in the production of four data cubes (two per filter associated with the detectors used). Each data cube posseses a pixel scale of 0.1\arcsec\ and spans specific wavelength ranges: 1.65 \micron\ -- 2.40 \micron, 2.48 \micron\ -- 3.16 \micron, 2.67 \micron\ -- 4.06 \micron, and 4.17 \micron\ -- 5.27 \micron, respectively.

\begin{table}[h]
  \centering
  \begin{tabular}{@{}ccc@{}}
    \hline
    \hline
    Parameter  & Value     \\
    \hline
    M$_{BH}$ & 10$^{9.5 \pm 0.2}$ M$_{\odot}$  \\
    L$_{bol}$ & 10$^{46.5 \pm 0.1}$ erg s$^{-1}$  \\
    L$_{Edd}$ &  10$^{47.5 \pm 0.7}$ erg s$^{-1}$ \\
    log $\lambda$$_{Edd}$ &  --1.0 $\pm$ 0.3 \\
    A$_{V,BAL}$ &   2.4 $\pm$ 0.3 mag \\
    A$_{V, SED}$ &   3.2 $\pm$ 0.8 mag \\
    \hline
  \end{tabular}
  \caption{F2M1106 Properties}
  
  \label{table:params}
\end{table}

\subsection{PSF Subtraction}
A significant challenge in studying extended emission in a quasar is the decomposition of the host galaxy from its central quasar. While JWST has high enough spatial resolution to enable studies of quasar hosts at higher redshifts, issues like surface brightness dimming and the quasar's bright PSF persist. The \qfit\ tool \citep{q3dfit23}, an adapted and extended iteration of \texttt{IFSFIT} \citep{rupk14, rupk17}, presents a promising avenue for effectively modeling and mitigating the influence of the quasar PSF, thereby unveiling the extended emission \citep{Wylezalek22, vayner23a, veil2023}. While Pa$\beta$, another line of interest, is detected near the nucleus (defined as an aperture with a radius of 0.5 kpc centered on the brightest pixel), we do not observe \feii\ in this nuclear region. Additionally, although Pa$\beta$ likely includes broad emission from the broad-line region, kinematic decomposition into broad and narrow components is not feasible given potential blending with unresolved narrow features. Thus, to preserve spatial structure, we begin with the non-subtracted cube and revisit PSF-subtracted measurements only where line ratios are critical for interpretation (i.e. Section \ref{subsec:shock}).

%therefore, in much of the subsequent \feii-focused analysis, we do not subtract a bright central source. The exception is Section \ref{subsec:shock}, where we analyze both \feii\ and Pa$\beta$ with and without PSF subtraction. 

We use q3dfit v1.1.4 for both cases. For the PSF-subtracted case, we identify the brightest spaxel as the central quasar and subtract it from the data. Each resulting spaxel is then fit with a model consisting of a third-order polynomial for the continuum and Gaussian profiles for the emission lines. The non-PSF subtracted case directly fits each spaxel with the same spectral model without any prior subtraction.  

%\textbf{While the Pa$\beta$ line is expected to include a broad component from the BLR due to the Type 1 AGN nature of the source, we initially analyze the non-PSF-subtracted data to preserve spatial information and avoid introducing artifacts from over-subtraction. Kinematic decomposition of the Pa$\beta$ line into distinct narrow and broad components is particularly challenging in this case, as the broad BLR emission may contain a partially unresolved narrow core that is difficult to isolate with our data. A clean separation would require assumptions about the line shape and higher signal-to-noise ratios than are available in most spaxels. Therefore, while we use the full observed Pa$\beta$ emission in our initial fitting across the field, we explicitly address potential contamination in regions where accurate line ratios are essential—namely, in Section \ref{subsec:shock}—by performing PSF subtraction and remeasuring key diagnostics.}

%Emission line analysis and maps
\subsection{[FeII] $\&$ Pa$\beta$ Morphology and Fitting}
\label{sec:morph_fit}

\begin{figure*}
    \centering
    \includegraphics[width=1.0\textwidth]{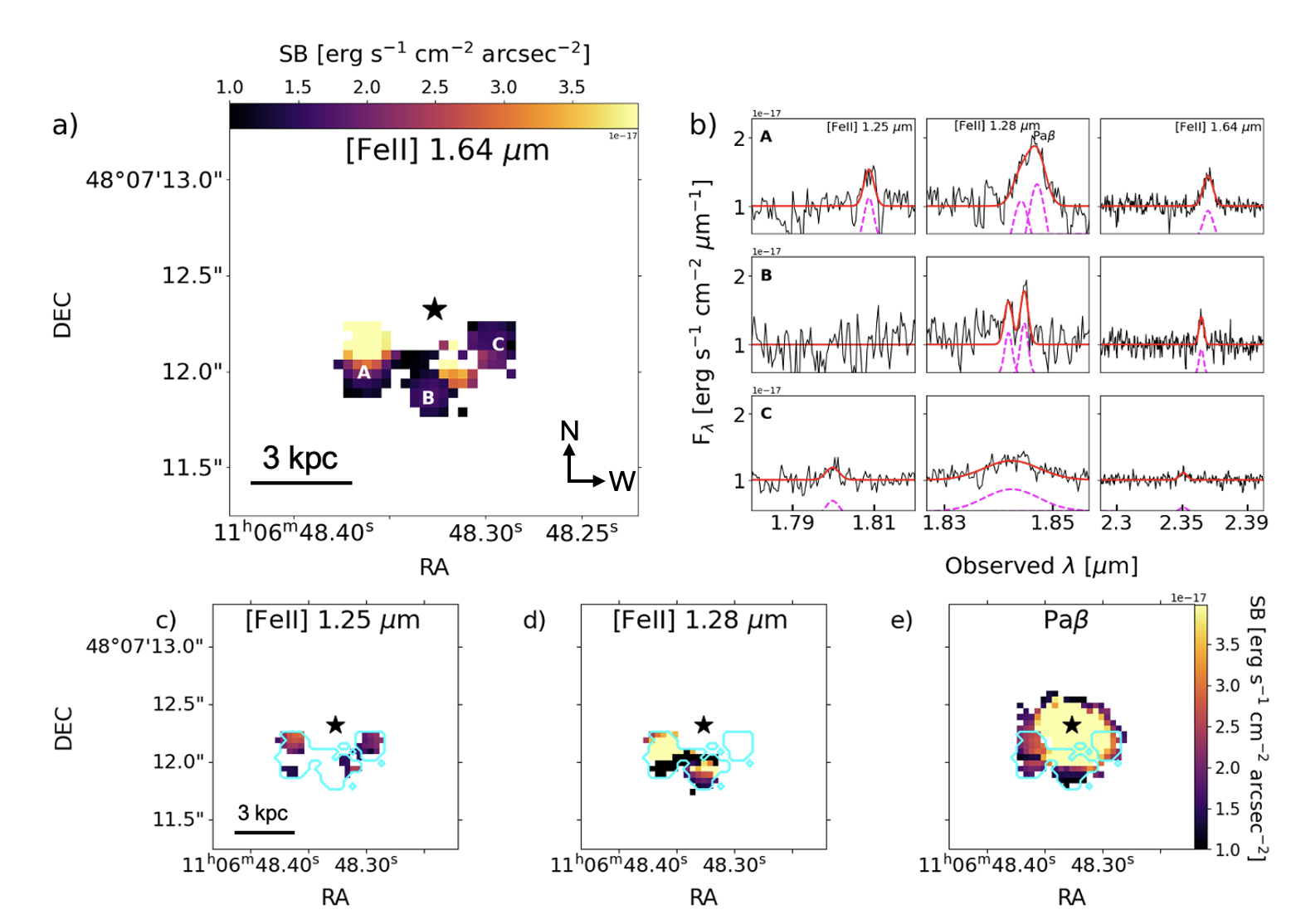}\\
    \caption{(a) JWST NIRSpec \feii\ 1.6440 \micron\ surface brightness map. The quasar is indicated by a black star, with representative spaxels A, B, and C selected. (b) Extracted spectra from these spaxels are shown in black, with Gaussian fits to the features in red and the individual Gaussian profiles in dotted magenta. The spectra include the \feii\ 1.2570, \feii\ 1.2791, \feii\ 1.6440 \micron, and Pa$\beta$ lines. For spaxel C, a second broad Gaussian component is present for the \feii\ 1.2791 \micron\ and Pa$\beta$ complex, but it has a lower flux amplitude and is cut off from the shown spectrum. (c, d, e) Surface brightness maps of the \feii\ 1.2570, \feii\ 1.2791 \micron, and total Pa$\beta$ (broad $+$ narrow component) lines, respectively, from left to right. These maps share the same spatial coordinates and color scale as the \feii\ 1.6440 \micron\ map in (a). Additionally, cyan contours in (b), (c), and (d) represent the \feii\ 1.6440 \micron\ emitting region for reference. All surface brightness maps include only spaxels with a signal-to-noise ratio (SNR) $>$ 2 in the respective lines.}
    
    \label{fig:linemaps}
\end{figure*}

\iffalse
\begin{table*}[htbp]
\centering
\begin{threeparttable}
\caption{\feii\ Emission Line Parameters}
\begin{tabular}{lcccccc}
\hline
\hline
Line & \multicolumn{2}{c}{Spaxel A} & \multicolumn{2}{c}{Spaxel B} & \multicolumn{2}{c}{Spaxel C} \\
     & $\lambda_c$ & $\sigma$ & $\lambda_c$ & $\sigma$ & $\lambda_c$ & $\sigma$ \\
\hline
\feii\ 1.25 $\mu$m & 1.80 & 194 & --- & --- & 1.79 & 260 \\
\feii\ 1.64 $\mu$m & 2.36 & 384 & 2.36 & 179 & 2.35 & 295 \\
\feii\ 1.28 $\mu$m & 1.84 & 384 & 1.83 & 179 & 1.83 & 295 \\
\hline
\end{tabular}
\begin{tablenotes}
\footnotesize
\item Centroid wavelengths ($\lambda_c$) are in $\mu$m, and velocity dispersions ($\sigma$) are in km\,s$^{-1}$.
\end{tablenotes}
\end{threeparttable}
\label{tab:feii_kinematics}
\end{table*}
\fi

\begin{table*}[htbp]
\centering
\begin{threeparttable}
\caption{\feii\ Emission Line Parameters}
\begin{tabular}{lcccccc}
\hline
\hline
Line & \multicolumn{2}{c}{Spaxel A} & \multicolumn{2}{c}{Spaxel B} & \multicolumn{2}{c}{Spaxel C} \\
     & $\lambda_c$ & $\sigma$ & $\lambda_c$ & $\sigma$ & $\lambda_c$ & $\sigma$ \\
\hline

\feii\ 1.25 $\mu$m 
    & $1.80$ & \B 194 $\pm$ \B 48 
    & --- & --- 
    & $1.79$ & \B 260 $\pm$ \B 91 \\

\feii\ 1.64 $\mu$m 
    & $2.36$ & \B 383 $\pm$ \B 27 
    & $2.36$ & \B 179 $\pm$ \B 33 
    & $2.35$ & \B 295 $\pm$ \B 34 \\

\feii\ 1.28 $\mu$m 
    & $1.84$ & \B 383 $\pm$ \B 27 
    & $1.83$ & \B 179 $\pm$ \B 33 
    & $1.83$ & \B 295 $\pm$ \B 34 \\

\hline
\end{tabular}
\begin{tablenotes}
\footnotesize
\item \textbf{Centroid wavelengths ($\lambda_c$) are in $\mu$m, and velocity dispersions ($\sigma$) are in km\,s$^{-1}$. Centroid wavelength errors are $\lesssim$ 0.0003 \micron\ and are ommitted for clarity.}
\end{tablenotes}
\end{threeparttable}
\label{tab:feii_kinematics}
\end{table*}

Figure \ref{fig:linemaps} presents examples of the emission line profiles for the \feii\ lines at 1.2570, 1.2791, and 1.6440 \micron, as well as for Pa$\beta$ (1.2822 \micron), taken from individual spaxels A, B, and C, along with their best-fit models in these representative regions. During fitting, we measure velocities relative to the host galaxy redshift, z = 0.4352.

The \feii\ lines, sharing the same ionization state of iron and similar excitation energies (7.87 eV), should exhibit shared kinematic properties and could be kinematically tied -- i.e., fit with the same velocity dispersion and centroids \citep{zaka16b}. In practice, the \feii\ 1.2570 \micron\ line shows a lower velocity dispersion than \feii\ 1.6440\micron, likely due to a lower signal-to-noise ratio (SNR) at that wavelength across the IFU cube. Specifically, the SNR for spaxels A, B, and C are approximately 5, 3, and 7 for \feii\ 1.2570 \micron, and approximately 11, 9, and 18 for \feii\ 1.6440 \micron. Additionally, \feii\ 1.2570 \micron\ is not detected to the south of the quasar (around spaxel B), where other \feii\ lines and Pa$\beta$ are present. Due to the contamination of the \feii\ 1.2791 \micron\ line in the Pa$\beta$ profile, which is otherwise the brightest in our spectra, we kinematically tie the \feii\ 1.2791 \micron\ and Pa$\beta$ lines to the \feii\ 1.6440 \micron\ line during fitting. The \feii\ 1.2570 \micron\ line, with its lower SNR and differing velocity dispersion, was fit separately. The measured centroids and velocity dispersions for the different \feii\ lines in spaxels A, B, and C are summarized in Table \ref{tab:feii_kinematics}. We find a single Gaussian fit to be sufficient for these lines.

Figure \ref{fig:linemaps} presents the surface brightness maps produced by \qfit\ for the emission features. From these maps, we observe that there is a clumpy morphology of the \feii\ 1.6440 \micron\ emission concentrated to the south of the quasar within the NIRSpec field of view. The Pa$\beta$ map, in contrast, shows a continuous distribution, with the central part dominated by the broad nuclear emission and narrower lines in areas where we detect \feii. The \feii\ 1.2791 \micron\ emission also appears clumpy, concentrated to the southeast and the south of the quasar, although some emission to the southwest may be missed due to its blending with the broad Pa$\beta$ profile. Additionally, the \feii\ 1.2570 \micron\ emission is observed in clumps to the southwest and southeast of the quasar. All surface brightness maps include only spaxels with a signal-to-noise ratio (SNR) $>$ 2 in the respective lines. The noise level, derived from the variance cube, has a relative standard deviation of $\sim$55$\%$. However, we find no spatial correlation between the observed clumpy structures and regions of higher or lower noise, suggesting that the clumpiness is not driven by noise variations.

%In contrast, the Pa$\beta$ emission extends northward, including and beyond the nuclear region.

\subsection{Comparison with the \oiii\ Outflow}
\label{subsec:kinematics}
%feii_oiii_comparison.png

%kin_comp_withoiii
\begin{figure*}
    \centering
    \includegraphics[width=0.9\textwidth]{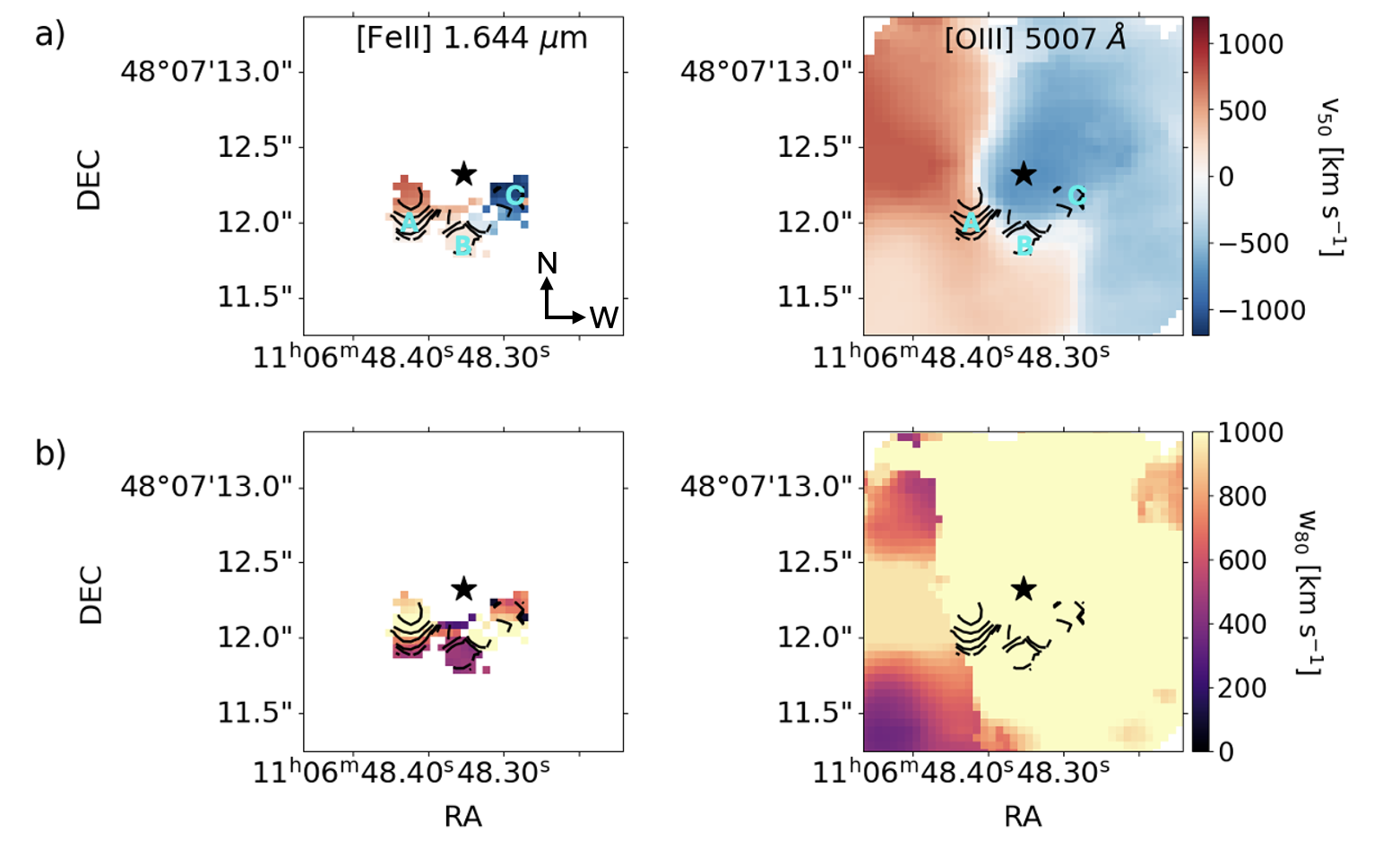}\\
    \caption{(\textit{row (a), left}) v$_{50}$ map of JWST NIRSpec \feii\ 1.6440  \micron. (\textit{row (a), right}) v$_{50}$ map of Gemini GMOS \oiii\ $\lambda$5007 \AA. Row (b) similarly plots for the W$_{80}$ maps of \feii\ (left) and \oiii\ in \kms\ (right) respectively. We overplot a \feii\ 1.6440  \micron\ surface brightness contour on all four subplots with levels $1.0$, $1.4$, $1.95$, $3.0$, $4.5$ and $5.0 \times 10^{-17}$~erg\,s$^{-1}$\,cm$^{-2}$\,arcsec$^{-2}$ for spatial comparison. Spaxels A, B, and C are labeled on the first row subplots.}
    \label{fig:oiiioverlay}
\end{figure*}

The outflow in F2M1106 was previously detected via \oiii$\lambda$5007 \AA\ by \citet{shen23}. Here, we compare the properties of \feii\ 1.6440 \micron\ and \oiii\ as illustrated in Figure \ref{fig:oiiioverlay} to determine if \feii\ is associated with the outflow.

Figure \ref{fig:oiiioverlay} reveals that the velocities of the \feii\ gas consist of two major components. One component is a redshifted nebula in the southeast of the NIRSpec field of view with a median velocity ranging from $\sim$50 \kms\ to 1100 \kms. The other component is a blueshifted nebula towards the southwest with v$_{50}$ ranging from around $\sim-$1200 \kms\ to 0 \kms. These high velocities exceed the gravitational potential limits, as galaxy rotation velocities typically remain within a few hundred \kms\ \citep{reun2002, reun2003}. This discrepancy between the velocity of ionized gas and the rotational velocity of the host galaxy is further supported by observations of the stellar component of the host galaxy (Rupke et al., in prep), which exhibits a rotational field with inverse velocity gradients (i.e., blueshifted to the east and redshifted to the west). 

Additionally, the velocity values (v$_{50}$) and the orientation of the \feii\ closely match the \oiii\ outflow. Therefore, we conclude that a majority of the \feii\ is part of the outflow and not related to the galaxy's rotation. Figure \ref{fig:oiiioverlay} also presents a comparison of the \feii\ and \oiii\ W$_{80}$ values, which represent the velocity width encompassing 80$\%$ of the total flux. While this parameter closely resembles the FWHM for a purely Gaussian profile, W$_{80}$ is more sensitive to the broad bases of non-Gaussian emission line profiles such as wings, making it suitable for analyzing high-velocity motions, including outflows and outflow-impacted emission \citep{liu13b, zaka14, harr14}. 

Overall, most regions of the nebula show similar W$_{80}$ values, with both \feii\ and \oiii\ exhibiting high W$_{80}$ velocities $>$ 1000 \kms\ (seen as a bright yellow ``band" in the \oiii). \citet{shen23} interpret this ``band" as resulting from the overlap of the receding and approaching sides of the outflow, as indicated by the presence of both redshifted and blueshifted components in this region in the v$_{50}$ map. This is consistent with the findings of \citet{liu24}, who observe localized regions of high dispersion in Pa$\alpha$—another outflow tracer—at the spatial resolution of the JWST observations. However, some notable differences between \feii\ and \oiii\ kinematics emerge. Not all spaxels for \feii\ exhibit high W$_{80}$ values within this region. Specifically, towards the south and southeast, the \feii\ component in a few spaxels appears narrower compared to the \oiii\ ($\sim$460 \kms\ for the \feii\ vs $\sim$1630 \kms\ for the \oiii\ in the south and $\sim$590 \kms\ vs $\sim$1600 \kms\ in the southeast). The \feii\ emission in the south, in particular, exhibits lower velocity dispersion and a net median velocity of a few hundred \kms, potentially indicating that unlike emission to the southeast and southwest, it is not associated with the outflow, but rather located within the host galaxy.

\subsection{\feii\ Extinction } \label{sec:extinction}

\begin{figure*}
    \centering
    \includegraphics[width=\textwidth]{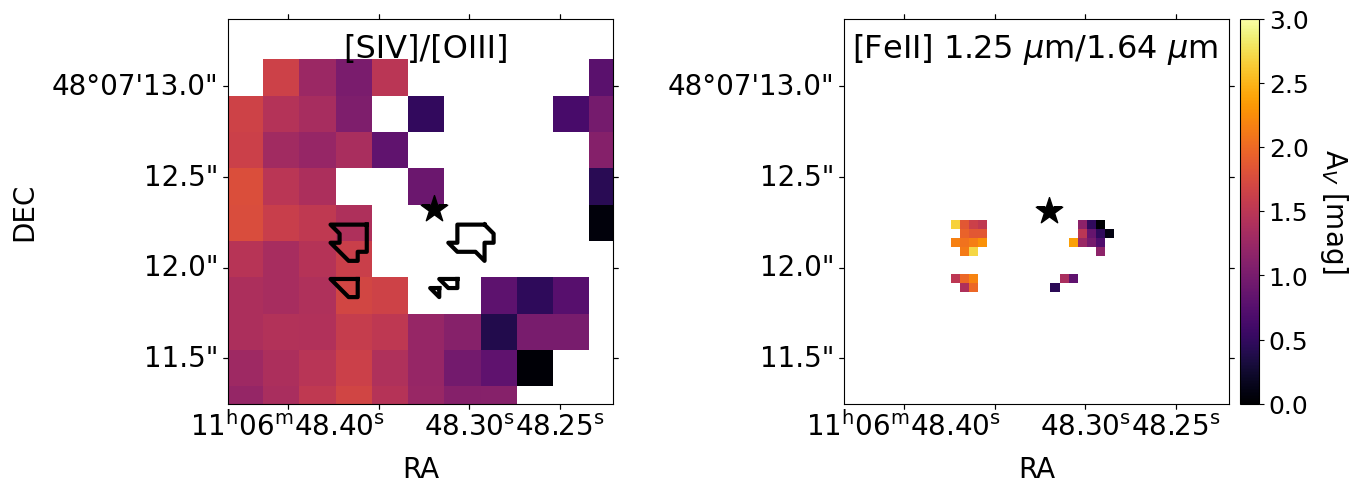}\\
    \caption{(\textit{left}) Visual extinction A$_{V}$ derived from the \oiii/[SIV] ratio, as presented in \citet{rupk23}, overlaid with an outline (in black) of the \feii\ 1.2570/1.6440 \micron\ extinction map distribution. (\textit{right}) A$_{V}$ map from the \feii\ 1.2570/1.6440 \micron\ ratio.}
    \label{fig:reddening}
\end{figure*}

We determine the extinction by analyzing the relative flux between \feii\ lines at 1.2570 and 1.6440 \micron, both originating from the same upper state (a$^{4}$D$_{7/2}$) \citep{anton2014, erkal2021}. This ensures that their intrinsic ratio is governed solely by their respective Einstein A coefficients and wavelengths, specifically: $(I_{1.25}/I_{1.64})_{0}$$=$ (A$_{ki,1.26}$/A$_{ki,1.64}$)/(1.26/1.64). Previous studies \citep{card89, gian2015, koolee2015} have determined this ratio theoretically and observationally, reporting values ranging from 0.94 to 1.49. For our calculations, we adopt the more commonly quoted theoretical ratio of $(I_{1.25}/I_{1.64})_{0}$$=$ 1.36 from \citet{card89}. Extinction is then determined using the following formula, where E$_{B-V}$ $=$ 0.3 $\times$ E$_{J-H}$ \citep{drai89} and A$_{V}$ $=$ 3.1 $\times$ E$_{B-V}$:

\begin{eqnarray}
I_{1.25}/I_{1.64} = (I_{1.25}/I_{1.64})_{0} \times 10^{-(E_{J-H})/2.5}.
\end{eqnarray}

The \feii\ extinction (A$_{V}$) map is shown in Figure \ref{fig:reddening}. Here, only spaxels with emission of both \feii\ 1.2570 and 1.6440 \micron\ larger than 3$\sigma$ are included, with the measurements mostly confined to the southwest and southeast of the quasar. For comparison, Figure \ref{fig:reddening} also shows the A$_{V}$ map from \citet{rupk23}, where dust extinction is estimated by comparing the observed [SIV]/\oiii\ line ratios. Similarly to their result, we find that the southeast \feii\ clump experiences higher extinction (median A$_{V}$$\sim$2.5 mag from the \feii\ and A$_{V}$$\sim$2.0 mag from [SIV]/\oiii\ map) in comparison to the southwest one, where the extinction is relatively low (median A$_{V}$$\sim$0.1 mag from the \feii\ and A$_{V}$$\sim$0.3 mag in [SIV]/\oiii). \citet{rupk23} attribute their observations to the bipolar outflow model, where the redshifted region experiences greater dust attenuation due to obscuration by the host galaxy, while the blueshifted region is less obscured and shows less attenuation. 

While both datasets qualitatively suggest higher extinction of the redshifted side, the specific measured values of A$_{V}$ differ somewhat. This is due to significant uncertainties, as both methods use intrinsic ratios that vary in the literature. Additionally, the extinction derived from \feii\ experiences systematic errors due to the differing velocity dispersions between the two lines. In our extinction measurements, we have not kinematically tied \feii\ 1.2570 and 1.6440 \micron\ fits, and therefore if the \feii\ 1.2570\micron\ velocity width is under-estimated because of its low SNR as discussed in Sec. \ref{sec:morph_fit}, then its flux is under-estimated and the extinction is under-estimated. The [SIV]/\oiii\ extinction measure faces uncertainties from the depletion fraction of sulfur. Furthermore, \feii\ and [SIV]/\oiii\ have different ionization energies (7.87 eV for \feii\ and 35.1 eV for [SIV] and \oiii) and thus trace different gas phases.

\subsection{Physical Conditions}

\begin{figure}
    \centering
    \includegraphics[width=0.49\textwidth]{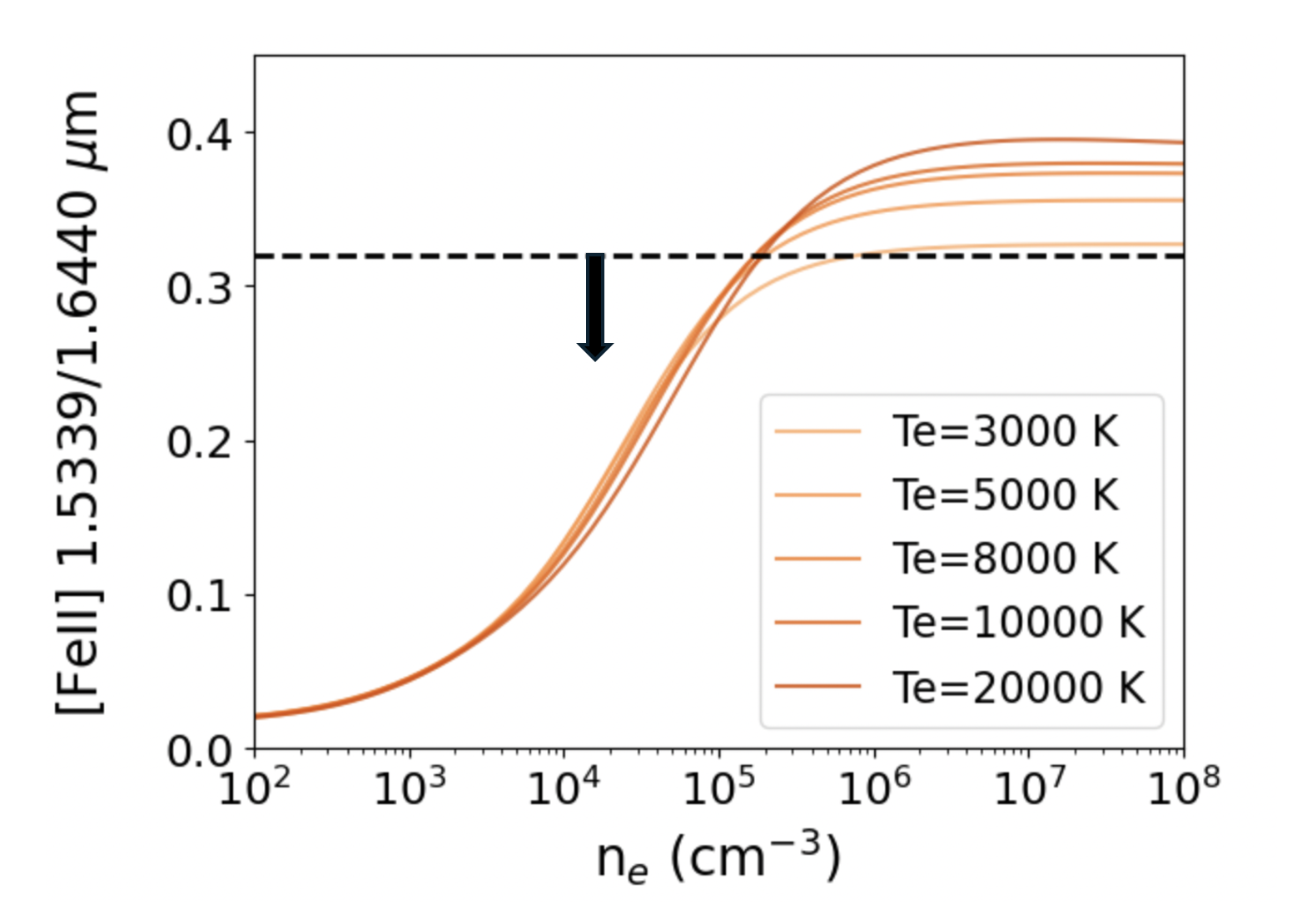}\\
    \caption{The model [Fe II] emission-line ratios versus electron density for the ratio [FeII] 1.5339/1.6440 $\mu$m. Each sequence of models correspond to a different temperature – from bottom to top – 3, 5, 10 and 2 x 10$^{3}$ K. We compute a 3$\sigma$ upper limit for the ratio in spaxels containing \feii\ 1.6440 \micron. The average value of the ratio is taken as an upper limit for the density, n$_{e}$$<$10$^{5.3}$\eden.}
    \label{fig:densityDiag}
\end{figure}

To determine the electron density of the \feii\ emission, we utilize the density diagnostic \feii\ 1.5339/1.6440 \micron\ line ratio. Given that the \feii\ 1.5339 \micron\ line is not detected in our spectra, we establish a 3$\sigma$ upper limit on its flux while calculating the average \feii\ 1.5339/1.6440 \micron\ ratio for spaxels containing \feii\ 1.6440 \micron\ emission. Employing the \pyneb\footnote{\url{http://research.iac.es/proyecto/PyNeb/}} package \citep{Luridiana15}, we compute electron densities for temperatures ranging from 3000 K to 20000 K, as depicted in Figure \ref{fig:densityDiag}. Our analysis yields an upper limit on the electron density of n$_{e}<$10$^{5.3}$ \eden. Studies using \feii\ density diagnostics typically report such high densities, as the critical density for \feii\ ranges from 10$^{4}$ to 10$^{5.5}$ \eden, making \feii\ a tracer for regions that are typically somewhat denser than those traced by \oiii. Specifically, \citet{strochi2009} measure densities ranging from 10$^{4}$ to 10$^{5}$ \eden\ in the nuclear region of NGC 4151 using the \feii\ 1.5339/1.6440 \micron\ line ratio, tapering off to 10$^{3}$ \eden\ beyond the central 0.5''. Similarly, \citet{mazz2006} find n$_{e}\sim$10$^{5.2}$ \eden\ for the nuclear region in Mrk 1210.

\section{Discussion}
\label{sec:discussion}

\subsection{Role of Radio Jets in \feii\ Ionization}
\label{subsec:radiosup}

\begin{figure}
    \centering
    \includegraphics[width=0.49\textwidth]{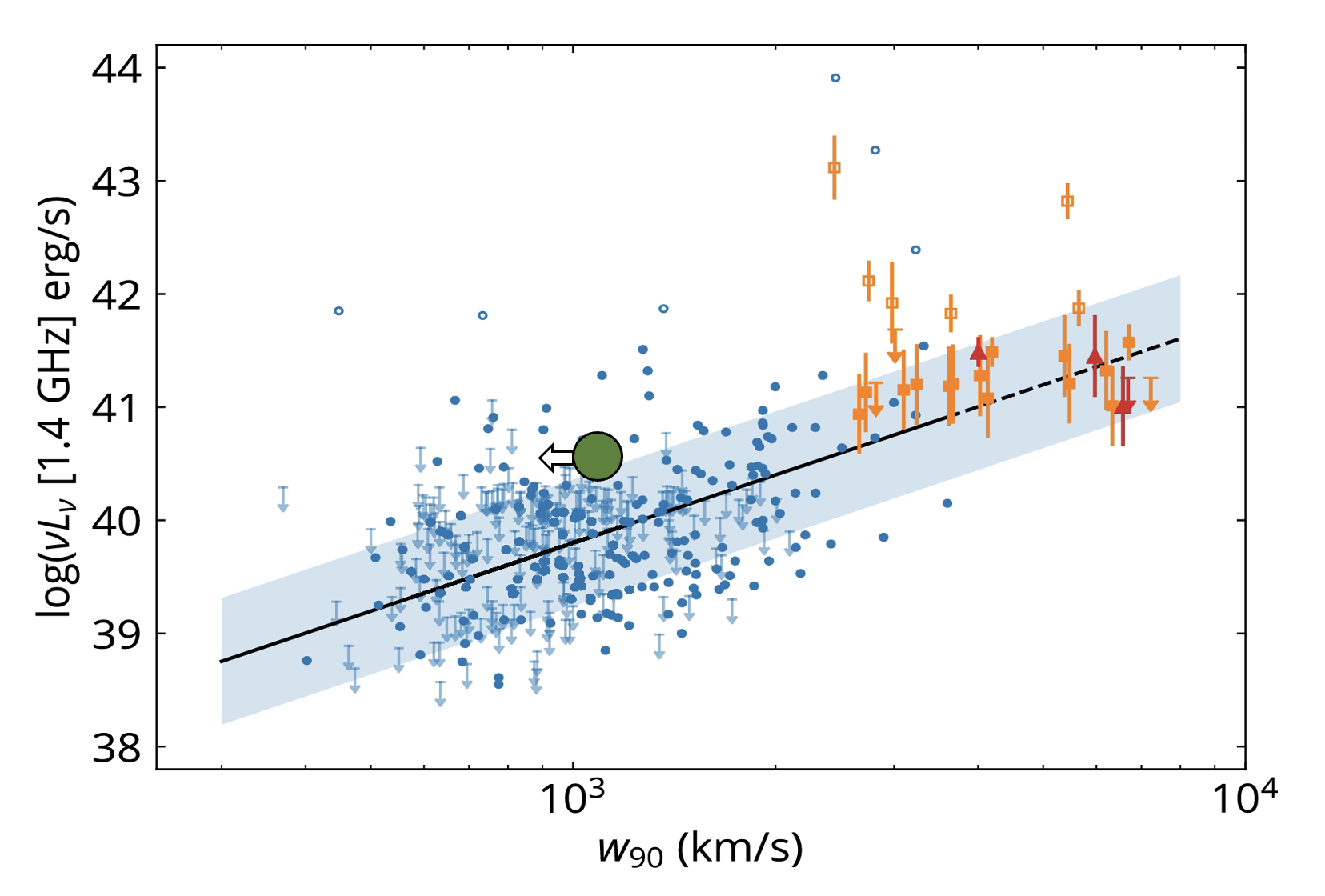}\\
    \caption{Radio luminosity at 1.4 GHz versus the velocity width W$_{90}$ of the \oiii\ emission for a radio-quiet sample, adapted from \citet{huang2018}. The blue dots denote low-redshift (z $<$ 0.8) type 2 quasars from \citet{zaka14}, with radio luminosity upper limits around 5$\sigma$ from FIRST and NVSS. The red triangles represent data from \citet{zaka16a}, and the orange squares are from \citet{perr19}, both depicting an extremely red quasar population, with \oiii\ observations at redshifts near z$\sim$3, with radio luminosity upper limits of 4$\sigma$. The black line illustrates the quadratic fit derived solely from the low-redshift radio-quiet sample (blue dots), with the shaded region indicating the root-mean-square scatter. Open symbols mark candidate radio-loud sources. F2M1106 (this study) is shown as a green circle.}
    \label{fig:radio_oiii}
\end{figure}

Our target is radio-quiet and point-like in the radio, with a steep radio spectrum ($\alpha$ $=$ -0.88). Typically, radio emission in active galactic nuclei is associated with processes such as jet activity. However, \citet{zaka14} study outflow velocites and radio luminosities of 568 obscured luminous quasars and find them to be strongly correlated in radio-quiet sources within their sample. The scenario they propose suggests that the radio emission observed in radio-quiet quasars is due to relativistic particles accelerated in the shocks within the quasar-driven outflows. More recently, \citet{huang2018} find the radio luminosity of their sample, which consists of 108 extremely red quasars at redshifts $z = 2–4$, to be related to the velocity dispersion of the \oiii\-emitting ionized gas, drawing similar conclusions. 

% talk a bit more about the correlation and what it means
We take the highest velocity dispersion W$_{90}$ value for the \oiii\ as an upper limit and place F2M1106 on the radio luminosity vs \oiii\ relation in Figure \ref{fig:radio_oiii} \citep{huang2018}. F2M1106 lies close to the \oiii-radio emission correlation of radio-quiet quasars. We find no evidence for a powerful radio jet in F2M1106, and therefore the observed \feii\ is unlikely to be jet-excited. While there could still be a small nuclear jet in the object, it is unlikely to be responsible for the \feii\ clumps.

\subsection{Supernovae-Driven Shock Ionization}
\label{subsec:supernovashock}

In Figure \ref{fig:oiiioverlay}, we demonstrate that the kinematics of \feii\ and \oiii\ are largely consistent with each other. Therefore, most of \feii\ is embedded in the quasar-driven outflow, with the possible exception of the southern clump of \feii\ which shows very low velocity dispersion compared to the values seen in \oiii. Here we examine whether supernova-driven shocks in the host galaxy would be sufficient to reproduce any of the observed \feii\ fluxes.

To calculate the supernova rate required to produce the observed \feii\ emission, we use the empirical relationship between the supernovae (SN) rate and \feii\ 1.2570 \micron\ line for starburst galaxies derived by \citet{rose12}:

\begin{equation}
  \begin{aligned}
      & \log\frac{\nu_{SN rate, \feii_{1.26} \micron}}{\rm{yr}^{-1}} =\\
      & (1.01 \pm 0.2)\log\frac{L(\feii_{1.26\micron})}{\mbox{erg s}^{-1}} - (41.17 \pm 0.9)
      \label{eq:snr}
  \end{aligned}
\end{equation}.

Due to the poor signal-to-noise ratio of the \feii\ 1.2570 \micron\ line in our observations and its nondetection to the south of the quasar, we use the \feii\ 1.6440 \micron\ line for our analysis instead, accounting for extinction where applicable. We select three apertures centered on spaxels A, B, and C, with diameters of 4.3'', 2.6'', and 2.5'', respectively, and determine the average extinction-corrected \feii\ 1.6440 \micron\ flux per aperture. These aperture sizes were chosen based on the spatial regions where the \feii\ signal is significant, as identified from the flux distribution and signal-to-noise ratio maps. The integrated luminosities of \feii\ 1.6440 \micron\ in each aperture are 5.5 $\times$ 10$^{41}$ erg s$^{-1}$, 3.7 $\times$ 10$^{40}$ erg s$^{-1}$, and 5.9 $\times$ 10$^{40}$ erg s$^{-1}$ respectively. Using these values in equation \ref{eq:snr}, we compute integrated SN rates of $\nu_{\text{SNrate, \feii}} \sim$10, 0.6, and 0.9 yr$^{-1}$ per aperture. The star formation rate upper limit estimated in Section \ref{subsec:sed_sec} of $\sim$132 M$_{\odot}$ yr$^{-1}$ corresponds to the supernova rate $\sim$2.2 yr$^{-1}$ \citep{kenn98}, assuming the Salpeter initial mass function in the mass range of 0.1–100 M$_{\odot}$ and solar metallicity. Therefore, the observed \feii\ emission of aperture A cannot be explained by star formation alone and requires additional sources of ionization. While the supernova rate from star formation alone exceeds the rate required to produce the \feii\ emission in apertures B and C, all of the star formation would have to be concentrated in these clumps to explain their \feii\ emission. While this scenario could plausibly result from positive feedback from the \oiii\ outflow—compressing the interstellar medium and triggering localized star formation—this interpretation should be approached with caution given the uncertainties in the star formation rate estimate, which is an upper limit and relies on the assumptions of the SED fit in Section \ref{subsec:sed_sec}.

%Additionally, given there is a wide range of galaxy templates that are in principle consistent with the SED in Section \ref{subsec:sed_sec}, the star formation rate estimate itself should be taken as approximate.

\subsection{MAPPINGS III Shock Models}
\label{subsec:shock}

%(i.e. the temperature of the shocked region)

\begin{figure}
    \centering
    \includegraphics[width=0.45\textwidth]{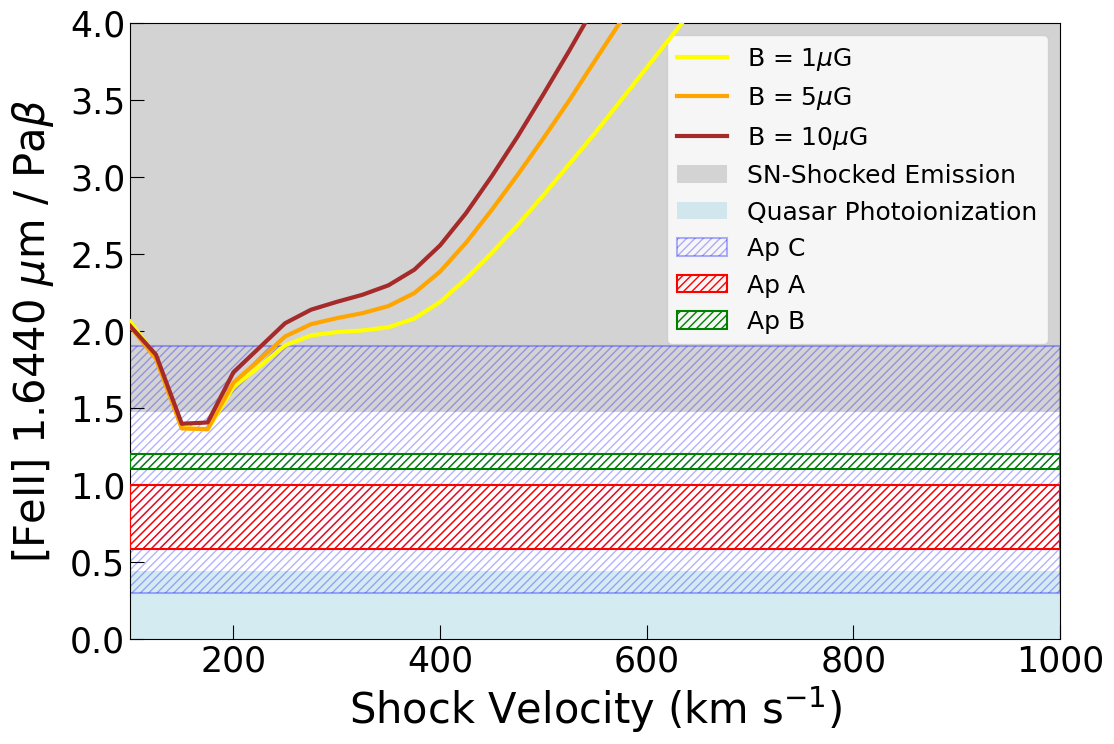}\\
\caption{MAPPINGS III shock diagnostics using the kinematic components of \feii\ 1.6440 \micron\ and narrow Pa$\beta$ from fitting. The shock+precursor models are taken from \cite{alle08} for an electron density of 1000 \eden and a solar metallicity. We allow the magnetic field to vary between values of 1, 5, and 10 $\mu$G cm$^{3/2}$. Additional line ratio ranges expected from pure supernovae-shock ionization and pure photoionization are indicated in light gray and light blue respectively \citep{rodri2004}. For each aperture (A: red, B: green, C: blue), shaded bands represent the range of extinction-corrected, mean integrated \feii/Pa$\beta$ ratios obtained using two methods—non-PSF-subtracted and PSF-subtracted Pa$\beta$—which we interpret as lower and upper bounds, respectively.}
\label{fig:shockmodels}
\end{figure}

We use MAPPINGS III \citep{alle08} models to determine the source of ionization of \feii. These models are particularly relevant for fast shocks, where the ionizing radiation arises from the cooling of hot gas heated by shocks. This process generates a substantial field of extreme ultraviolet and soft X-ray photons, accounting for ionization from both quasar outflow-driven shocks and supernovae-driven shocks. The models include the following parameters: shock velocity (ranging from 100 to 1000 km s$^{-1}$ in increments of 25 km s$^{-1}$), preshock density (ranging from 10$^{-2}$ to 10$^{3}$ cm$^{-3}$ in logarithmic steps of 10), and preshock transverse magnetic field strength (ranging from 10$^{-4}$ to 10$^{3}$ $\mu G$).

The \feii\ 1.2570 \micron/Pa$\beta$ line ratio is a well-established diagnostic for shocked emission. This ratio typically falls below 0.6 in starbursts, but rises above 2.0 in regions where shocks driven by supernovae are the primary mechanism \citep{rodri2004}. In active galaxies, the ratio generally ranges between 0.6 and 2.0, with values near 0.6 indicating photoionization (either from the active galactic nuclei or ongoing star formation) as the prevailing process, while values approaching or exceeding 2.0 suggest a greater contribution from shock excitation, particularly supernovae-driven \citep{mour00, rodri2004, riff08, strochi2009}. 

Due to the low signal-to-noise ratio of the \feii\ 1.2570 \micron\ line in our data, we instead use the \feii\ 1.6440 \micron/Pa$\beta$ line ratio as a proxy. A complicating factor is that Pa$\beta$ contains a strong broad component near the nucleus, which can contaminate the extended emission and bias line ratio measurements in adjacent regions. Moreover, due to poor signal-to-noise, a robust decomposition of the Pa$\beta$ line into narrow and broad components is not feasible across the full field of view. We therefore adopt two complementary approaches: (1) using the non-PSF-subtracted Pa$\beta$ map, and (2) subtracting a modeled PSF of the broad Pa$\beta$ emission before measuring ratios. While method (1) may underestimate the true \feii/Pa$\beta$ ratio due to contamination by the broad-line component, method (2) may slightly overcorrect by removing some extended narrow-line flux.

%However, due to the low signal-to-noise of the \feii\ 1.2570 \micron\ line, we instead use the \feii\ 1.6440 \micron/Pa$\beta$ line ratio for our analysis. An additional challenge here is that Pa$\beta$ contains a strong broad component near the nucleus, which can contaminate measurements in adjacent regions. In addition, decomposition of Pa$\beta$ into narrow and broad components is complicated by the poor signal-to-noise ratio. To address this, we take two complementary approaches: 1) we use the non-PSF-subtracted Pa$\beta$ map, and 2) we PSF-subtract the Pa$\beta$ broad component before measuring ratios. While the former approach may underestimate the \feii/Pa$\beta$ ratio in regions where broad-line flux spills over, the latter may slightly overcorrect, potentially removing some extended narrow-line emission.

We treat these two methods as representing lower and upper bounds on the intrinsic \feii/Pa$\beta$ ratio in each region. Figure \ref{fig:shockmodels} displays the extinction-corrected, mean integrated \feii\ 1.6440 \micron/Pa$\beta$ line ratios for apertures A, B, and C, with shaded bands representing the range spanned by the two measurement methods (non-PSF-subtracted and PSF-subtracted Pa$\beta$), which we interpret as lower and upper bounds, respectively. We compare these values to the shock models, which, while useful for illustrative purposes, assume a simplified one-zone geometry and therefore may not capture the full complexity of ionization structures in this system. Since the models are largely insensitive to variations in electron density, we fix the electron density at 1000 cm$^{-3}$. We also assume solar metallicity for both the radiative shock and precursor components and allow the magnetic field strength to vary from 1, 5, and 10 $\mu$G cm$^{3/2}$. Additionally, we highlight line ratio ranges expected for pure supernova shock excitation and purely photoionized regions, converted to \feii\ 1.6440 \micron/Pa$\beta$ ratios using an intrinsic line flux ratio of $(I_{1.25}/I_{1.64})_{0} = 1.36$ \citep{card89}.

We find that the line ratio ranges for apertures A and B fall into an intermediate regime: aperture A ranges from 0.58 to 1.0, consistent with a mix of photoionization and low-level shocks, while aperture B spans a narrower range (1.1–1.2), overlapping the lower end of shock-dominated scenarios. Aperture C shows the widest variation (0.3–1.9), straddling nearly the entire diagnostic range and underscoring the difficulty of unambiguously classifying the dominant ionization mechanism there. In particular, while the upper limit in aperture C approaches the threshold expected for shocks, the lower bound is more consistent with photoionization by the quasar.

%Given this ambiguity, and the limitations of the MAPPINGS models, we interpret the C aperture results with caution.

%Taken together, these results suggest that while shocks may contribute to \feii\ emission in apertures A and B--particularly in aperture B where the line ratio is consistently elevated—the dominant ionization source remains somewhat uncertain. The observed ratios are difficult to reconcile with pure supernova-driven shocks, especially given the unphysical star formation rates that would be required to explain the measured \feii\ luminosities 

Apertures A and B have a well-determined \feii/Pa$\beta$ line ratio which is inconsistent with pure photoionization from the quasar. In particular, aperture B shows the highest value of the line ratio, and it is likely that shocks contribute to \feii\ ionization in this region. But the measured \feii\ luminosities are too high to attribute these shocks to star formation (as discussed in Section \ref{subsec:supernovashock}). In the absence of strong evidence for concentrated star formation, quasar-driven outflows remain a plausible mechanism for the observed shock signatures in apertures A and B, either directly through mechanical interaction or indirectly through pressure-driven compression and induced star formation.

In aperture B, spatial and kinematic separation between \feii\ and \oiii\ supports the idea that the outflow is interacting with clumpy or offset ISM structures rather than uniformly permeating the host. Theoretical models of wide-angle quasar winds support this scenario, suggesting that outflows can compress cold gas and generate local shocks even outside the regions directly illuminated by the AGN \citep{wagn13, king2011, nayakshin2012}. Still, given the range of line ratios and the challenges in accurately disentangling narrow and broad Pa$\beta$ emission, the inferred shock contribution should be treated as suggestive rather than conclusive.

%This adjustment increases the ratios to 1.0, 1.2, and 1.9, respectively, suggesting a greater contribution of shocks in all three apertures, which would have to be largely due to the quasar winds since the total amount of \feii\ emission in clumps A, B and C is inconsistent with star formation. However, given the risk of oversubtraction, we interpret these results with caution.

\section{Summary}
\label{sec:conclusions}

In this paper, we examine the ionizing mechanism of \feii\ emission in JWST NIRSpec IFU data of F2M1106, a red quasar at $z=0.4352$ with a known powerful outflow in \oiii. Our results are summarized as follows:
\begin{itemize}

\item
We model the spectral energy distribution (SED) of the object across optical to far-infrared wavelengths, incorporating data from GALEX, SDSS, 2MASS, WISE, and SOFIA. The SED reveals a continuum extinction of A$_{V} \sim$ 3.2 mag and a bolometric luminosity of L$_{bol}$ $=$ 10$^{46.5 \pm 0.1}$ erg s$^{-1}$. Analysis of the broad H$\alpha$ emission line indicates a black hole mass of M$_{BH}$ $\sim$ 10$^{9.5 \pm 0.2}$ M$_{\odot}$ and an Eddington ratio of log log$\lambda_{Edd}$ $\sim$ --1.0 $\pm$ 0.3.

%, suggesting the quasar operates near the Eddington limit

\item
Analysis of the the NIRSpec IFU data reveals evidence of \feii\ lines at 1.2570, 1.2791, and 1.6440 \micron\ and Pa$\beta$. All three \feii\ lines are clumpy, but not necessarily cospatial (\feii\ 1.2570 \micron\ is not detected in the southern region due to poor SNR; \feii\ 1.2791 \micron\ is blended with Pa$\beta$ in the southwest). The Pa$\beta$ map shows broad-line emission near the quasar, with narrower lines in regions where \feii\ is present. 

\item
We analyze the extinction in F2M1106 using the \feii\ 1.2570/1.6440 \micron\ line ratio, revealing higher extinction in the redshifted southeast region (A$_V \sim$ 2.5 mag) compared to the blueshifted southwest (A$_V \sim$ 0.1 mag). This is consistent with previous [SIV]/\oiii\ findings, but shows some variance due to method-specific uncertainties. The electron density is constrained using the \feii\ 1.5339/1.6440 \micron\ line ratio, with an upper limit of n$_{e}$ $<$ 10$^{5.3}$ \eden, indicating that the \feii-emitting regions are associated with dense gas.

\item Two of the three clumps of the \feii\ 1.6440 \micron\ emission are kinematically consistent with the outflow traced by \oiii$\lambda$5007 \AA. Both lines show two major velocity components: a redshifted nebula in the southeast (with \feii\ v$_{50}$ $\sim$50--1100 \kms) and a blueshifted nebula in the southwest (\feii\ v$_{50}$ $\sim$--1200--0 \kms), both with high velocity widths. These velocities, exceeding typical galaxy rotation speeds, align with the \oiii\ outflow. In contrast, the \feii\ emission in the southern region shows narrower velocity widths (W$_{80}$ $\sim$460 \kms) compared to \oiii\ (W$_{80}$ $\sim$1630 \kms), suggesting that part of the \feii\ emission in the south may be located within the host galaxy.

\item We use MAPPINGS III models as a diagnostic framework to contextualize the observed \feii/Pa$\beta$ line ratios across apertures A (southeast), B (south), and C (southwest). By bracketing the measurements with PSF-subtracted and non-subtracted Pa$\beta$ maps, we account for systematic uncertainties and interpret the resulting ratio ranges as indicative of mixed ionization mechanisms. Apertures A and B fall within or near the regime expected for partial shock ionization, while aperture C remains consistent with photoionization. We propose that quasar-driven outflows may be responsible for inducing shocks in A and B via compression or indirect propagation, though the simplified single-zone nature of the MAPPINGS models limits the strength of this conclusion.

%Using MAPPINGS III models, we find that the \feii/Pa$\beta$ line ratios indicate quasar photoionization as the primary ionization mechanism for \feii\ in apertures A (southeast \feii\ emitting region) and C (southwest). The higher ratio in aperture B (the south region) suggests partial shock ionization. We suggest that this region experiences increased back pressure from the quasar-driven outflow which exceeds the maximum host galaxy pressure and compresses the gas, inducing shocks.

\end{itemize}

\section{Appendix}

\begin{table}[h!]
    \centering
    \caption{F2M1106 Photometry}
    \begin{tabular}{lcc}
        \toprule
        Band & Wavelength ($\mu$m) & AB Mag \\
        \hline
        GALEX NUV  & 0.231 & 20.40 $\pm$ 0.06 \\
        SDSS u     & 0.355 & 19.81 $\pm$ 0.06 \\
        SDSS g     & 0.468 & 19.01 $\pm$ 0.03 \\
        SDSS r     & 0.616 & 18.25 $\pm$ 0.03 \\
        SDSS i     & 0.748 & 17.70 $\pm$ 0.03 \\
        SDSS z     & 0.893 & 17.37 $\pm$ 0.03 \\
        2MASS J    & 1.235 & 17.16 $\pm$ 0.10 \\
        2MASS H    & 1.662 & 16.51 $\pm$ 0.09 \\
        2MASS K    & 2.159 & 15.79 $\pm$ 0.06 \\
        WISE W1    & 3.352 & 14.34 $\pm$ 0.04 \\
        WISE W2    & 4.602 & 13.67 $\pm$ 0.04 \\
        WISE W3    & 11.560 & 12.78 $\pm$ 0.03 \\
        WISE W4    & 22.090 & 12.09 $\pm$ 0.04 \\
        SOFIA HAWC+ C & 62 & $<$ 14.89 \\
        SOFIA HAWC+ D & 107 & $<$ 14.65 \\
        \hline
    \end{tabular}
    \label{tab:photometry}
\end{table}

\begin{figure*}
    \centering
    \includegraphics[width=0.9\textwidth]{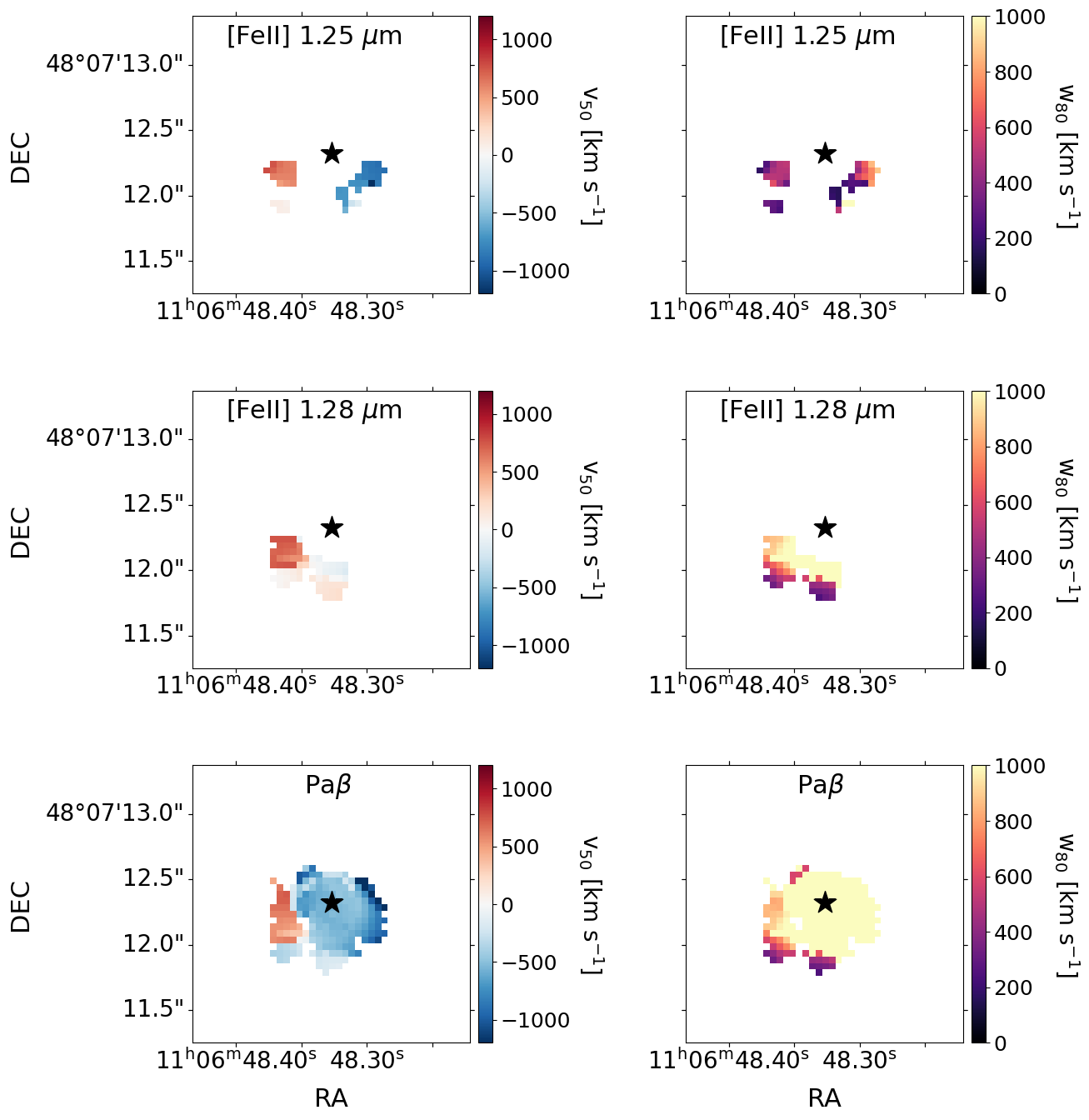}\\
    \caption{Velocity ($v_{50}$; left column) and velocity width ($W_{80}$; right column) maps derived from JWST/NIRSpec IFU observations for three near-infrared emission lines. From top to bottom, each row corresponds to the \feii\ 1.25\,$\mu$m, \feii\ 1.28\,$\mu$m, and (non-PSF subtracted) Pa$\beta$ emission. The location of the QSO is marked with a black star in each panel.}

    \label{fig:fe125_fe128_pab}
\end{figure*}

\begin{acknowledgments}
A.V., N.L.Z., Y.I. S.V, and S.S. are supported in part by NASA through STScI grant JWST-ERS-01335. S.S is also supported by STScI grant JWST-GO-01970. N.L.Z further acknowledges support by the Institute for Advanced Study through J. Robbert Oppenheimer Visiting Professorship and the Bershadsky Fund. The JWST data presented in this article were obtained from the Mikulski Archive for Space Telescopes (MAST) at the Space Telescope Science Institute. The specific observations analyzed can be accessed via \dataset[doi: 10.17909/da4v-6h61]{https://doi.org/10.17909/da4v-6h61}.

\end{acknowledgments}

%\facilities{}

%\software{astropy \citep{Astropy2013, Astropy2018},  
%          Cloudy \citep{Ferland2013}, 
%          }

%\appendix

%\section{Appendix information}

\bibliography{bib}{}
\bibliographystyle{aasjournal}

\end{document}